\documentclass{aa}
\usepackage{graphicx}

\begin{document}

%\def\und#1{$\underline{\smash{\hbox{#1}}}$}
%\def\alinp{\vskip 5 truemm \par \noindent}
%\def\alinppp{\vskip 2 truemm \par \noindent}
%\def\la{\mathrel{\hbox{\rlap{\hbox{\lower4pt\hbox{$\sim$}}}}}\hbox{$<$}}
%\def\ga{\mathrel{\hbox{\rlap{\hbox{\lower4pt\hbox{$\sim$}}}}}\hbox{$>$}}
%\def\np{\vfill\eject}
%\def\pan{\par\noindent}
%\settabs 11 \columns
%

% \thesaurus{06     
% A&A Section 6: Form. struct. and evolut. of stars
%              (19.63.1)} % Stars: structure of.
%
   \title{Mass transfer from the donor of GRS 1915+105}
%   \subtitle{      }

  \author{O. Vilhu
%         \inst
    }

   \offprints{O. Vilhu}

   \institute{
Observatory, Box 14, FIN-00014 University of Helsinki. Finland\\
      \email{osmi.vilhu@helsinki.fi} 
%\and 
%   SRON Laboratory for Space Research, Sorbonnelaan 2,
%               3584 CA Utrecht, The Netherlands\\
%              \email{R.Mewe@sron.nl}
%\and
%   Tuorla Observatory, University of Turku, Finland\\
%          \email{pahakala@astro.utu.fi}
                                           }

   \date{ accepted 16.4.2002 }

%   \maketitle

\abstract{ 
A scenario for a periodic filling and emptying of the accretion disc of
the microquasar GRS1915+105 is proposed, by estimating the mass transfer rate
from the donor and comparing it with the  observed accretion rate
onto the primary black hole. 
The mass of the Roche-lobe-filling donor (1.2 $\pm{0.2}$ M$_{\sun}$), 
the primary black hole mass ( 14 $\pm{4}$ M$_{\sun}$) and the binary orbital
period of 33.5 d  (Greiner et al. 2001b)
predict for the donor spectral type
and K-magnitude around K6 III and -2.6, respectively. 
The He-core of 0.28 M$_{\sun}$ of such a giant leads to evolutionary expansion along the
giant branch with a conservative mass transfer rate of 
  $\dot{M}_{d}$ =  (1.5 $\pm{0.5}$)  $\times$ 10$^{-8}$
M$_{\sun}$/year. On the other hand, the average observed 
 accretion rate onto the primary  
is  ten times larger:
 $\dot{M}_{obs}$ = 2.0 $\times$ ($\eta$/0.1)$^{-1}$(d/12.5kpc)$^2$10$^{-7}$ M$_{\sun}$/y, 
where $\eta$ is the efficiency of converting
accretion into radiation.
We propose a duty cycle with  (5-10)($\eta$/0.1) {\it  per cent} 
active ON-state. The timescale of the (recurrent) OFF-state  
is identified as the viscosity
time scale at the circularization radius (14 R$_{\sun}$) and equals 
t$_{visc}$  = 370 ($\alpha$/0.001)$^{-4/5}$   years, 
where $\alpha$ is the viscosity parameter in the $\alpha$-prescription
 of a classical disc. 
If the viscosity at the outer edge of the disc is small and $\eta$ is close to the maximum available
potential energy (per rest mass energy) at the innermost stable orbit,
the present activity phase may still last another 10 -- 20 years. We also
discuss other solutions allowing a broader range of donor masses 
(0.6 -- 2.4 M$_{\sun}$).     
                     }

        \maketitle
%________________________________________________________________

\section{Introduction}

Greiner et al. (2001a) identified the mass-donating secondary 
star of GRS 1915+105 to be
 a K-M III giant, indicating that this prototype microquasar is a low-mass
X-ray binary (LMXB). Further, using the 
Very Large Telescope (VLT) and the band-heads  of $^{12}$CO and
$^{13}$CO  , Greiner et al. (2001b) managed to obtain
 the radial velocity curve of the secondary. The orbital period of 33.5 days,
 the large mass function f(M) =
9.5 $\pm{3.0}$ M$_{\sun}$  and known jet-inclination (70$^{\degr}$) 
permitted to
constrain the primary black hole 
 mass between M$_{BH}$ = (10 -- 18) M$_{\sun}$, assuming
 the donor mass to lie between M$_d$ = (1.0 -- 1.4) M$_{\sun}$.
 The large BH mass
points to rapid rotation since the smallest inner disc radii modelled
(see e.g. Belloni et al. 1997; Vilhu et al. 2001) are as small 
as 20 km, close to the
last marginally stable orbit (0.5 R$_g$) of an extreme prograde Kerr-hole
of 14 M$_{\sun}$.

In the present paper we estimate the mass transfer rate from the evolving donor
but allowing a broader range  (0.6 -- 2.4) M$_{\sun}$ for its mass 
to include e.g.  a possible stripped giant. Further, we estimate the 
viscosity time scale
at the circularization radius and the amount of mass accumulated there.
Using the mean observed  accretion rate over the past 6 years (via luminosity
conversion) we  arrive at an estimate for the timescale of the 
possible duty cycle, relevant  also when discussing a link to
ultraluminous sources (ULX) in other galaxies (King et al. 2001).

\section{ Mass transfer rate from the evolving donor}

We assume that the donor fills its Roche lobe and that  the mass loss is
determined by  evolutionary expansion along the giant branch,
 conserving the orbital angular 
momentum. The properties  along the giant branch
(luminosity and radius)
depend mainly on the He-core mass and less on the envelope mass.
In this case an analytical simplification is possible 
(Webbink, Rappaport and Savonije, 1983) and the procedure is also  
 presented  by Verbunt and
van den Heuvel (1995). In particular, the radius and luminosity can be fitted
with 3rd order polynomials on the core mass.
%plot ______________________________________________
   \begin{figure*}
%     \vspace{4cm}
%\rule{0.4pt}{4cm}% line thickness, height of picture
\includegraphics{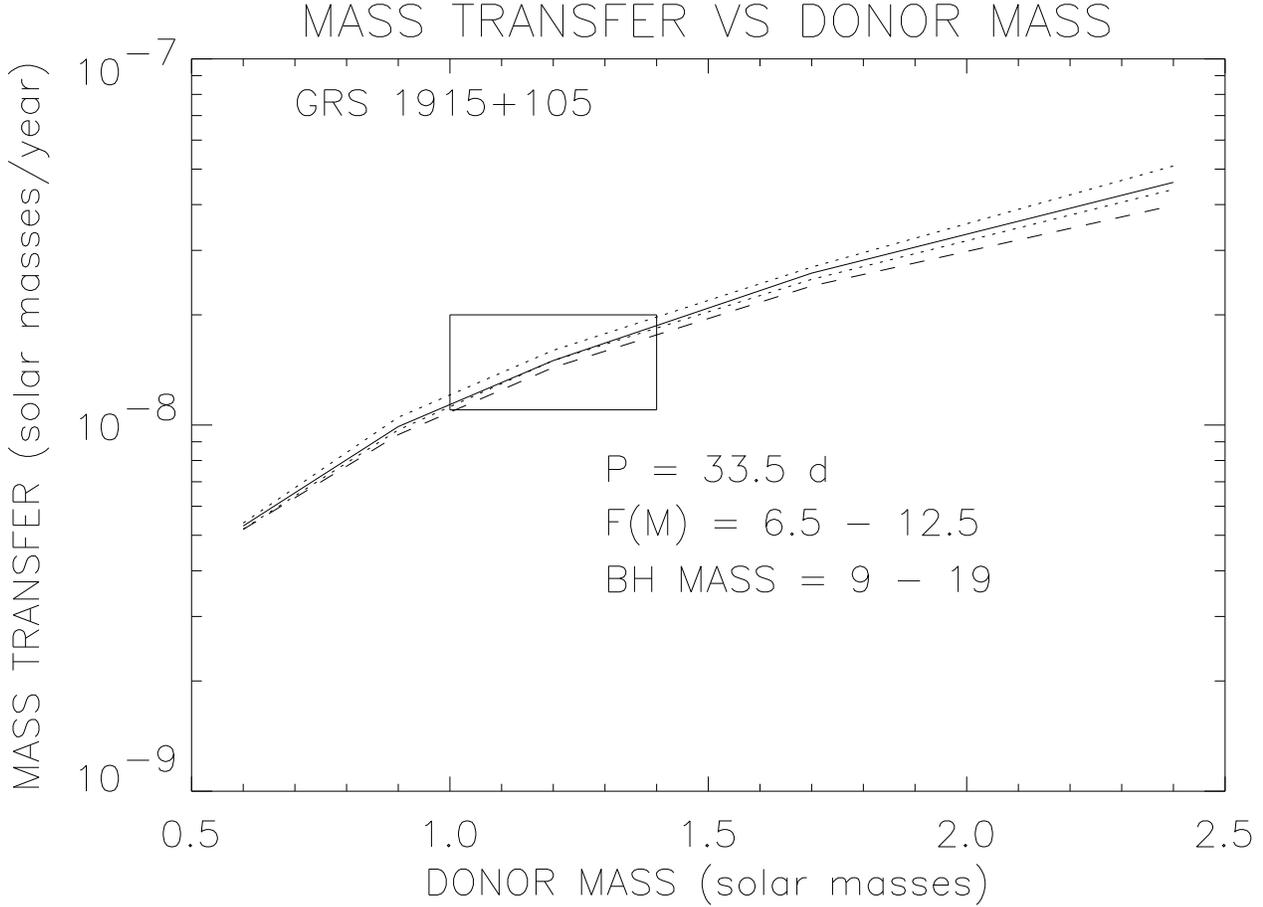}
%     \caption{ }
%\hfill      
\caption{  Conservative mass transfer rates 
 from the
evolving K-giant donor of GRS 1915+105 using the analytic 
methods by 
Webbink, Rappaport and Savonije (1983) and Pop. I abundances Z = 0.02.
 The curves were computed for
different donor masses (in M$_{\sun}$) using 
 three mass function  values inside the range 
f(M) = 9.5 $\pm{3.0}$ M$_{\sun}$ 
(Greiner at al. 2001b) with P$_{orb}$ fixed to 33.5 d (solid line:
f(M) = 9.5, dotted lines: f(M) = 6.5 and 12.5).  The dashed line  
represents the
analytic expression given by King et al. (2001, their eq. 12). The box shows the
mass range suggested by Greiner et al. (2001b). 
}
\label{Figdonor}
\end{figure*}

The growth of the core mass, resulting in an
 increase of the radius, is determined by the
luminosity due to hydrogen shell burning which, in turn, depends completely
on the core mass. In the conservative case,  fixing the binary
parameters and forcing the secondary to
fill its Roche lobe, it is rather simple to compute the core mass and 
consequently the 
mass loss from the donor (we use Pop. I abundances Z = 0.02;
for details see Webbink et al. (1983) and Verbunt and 
van den Heuvel (1995), p. 482).

The first line in Table 1 gives the results for the
best-fit masses given by Greiner et al. (2001b) 
(14 M$_{\sun}$ + 1.2 M$_{\sun}$, P=33.5 d).
The second line gives the parameter ranges  
 if the donor mass M$_{d}$ is varied between 
(0.6 -- 2.4) M$_{\sun}$ and satisfying
the mass function constraint f(M)=9.5$\pm{3.0}$ M$_{\sun}$.
The BH mass varies within these domains between 9--19M$_{\sun}$.
 The mass transfer rate depends mainly on the donor mass as shown
explicitly in Fig. 1.
 It follows closely the analytic expression given by King et al. 
(2001; their eq. 12, the dashed line in Fig. 1). In particular, the 
uncertainties in the mass function have minor effects on this relation.

\

\section{ Duty-cycle time scales}        
%\nopagenumbers
In the conservative case, the mass leaving the donor via the 
L1-point settles down into a Keplerian orbit around the primary BH, the radius of
which is called  the 'circularization radius'. This radius is  given in Table 1 
as computed from the  
analytic approximation to numerical data 
(Frank, King and Raine, 1992 (FKR), p. 56, eq. 4.18):

$$R_{circ} = 4(1+q)^{4/3}(0.500 - 0.227 LOG(q))^4P_{day}^{2/3},  \eqno(1)$$

%
%$$P_{jk}^{exc} = S_{exc}\ BR\ n_e\ N_{ion}, \eqno(2)$$
%
where q is the mass ratio M$_d$/M$_{BH}$.

Due to  the viscosity, the torus at R$_{circ}$ will be stretched and flattened 
into a disc  on a viscous time scale (by angular momentum transfer). The size of the  viscosity
is highly uncertain but in the $\alpha$-prescription of  classical disc theory
it is parameterized  and  the viscous time scale at R$_{circ}$ 
has a scaling law
(see FKR p. 99, eq. 5.63):

\small
$$t_{visc}  = 370 (\alpha/0.001)^{-4/5} (M_{BH}/14)^{1/4}  
(\dot{M}_d/1.5E-8)^{-3/10}$$

 $$\times(R_{circ}/14)^{5/4}   years  \eqno(2)$$

\normalsize
where the parameters are scaled to those used in Table 1 for the 
best-fit binary parameters and $\alpha$ = 0.001.
We may call this  the {\it recurrence time} during which a new disc is formed 
if the old
one has been  rapidly swallowed into the BH. 
The mass accumulated in the torus  
around R$_{circ}$ 
during this time equals  
to  M$_{accum}$ =  t$_{visc}$ $\times$ $\dot{M}_d$. 

Surprisingly, M$_{accum}$  is roughly equal to the mass  of a classical viscous disc 
(using $\alpha$-prescription, gas pressure and  Kramer's opacity)
if the outer radius is set equal to R$_{circ}$ and   
2$\times$10$^{-7}$ M$_{\sun}$/year 
is used for the disc accretion. 
This accretion rate can  be derived from
the RXTE observations over the past six years.  
The  ASM light curve gives a time-averaged mean value of 58 counts/s (0.77
in the Crab-units)
between 2 -- 13 keV which 
corresponds to a total intrinsic luminosity 
L = 1.2 $\times$ 10$^{39}$ (d/12.5 kpc)$^2$ erg/s  using PCA+HEXTE fits by 
Vilhu et al. (2001). 
 The distance d is scaled to the mean value 12.5 kpc given by Chaty et al. (1996) with $\pm{1.5}$ kpc uncertainty.     
 This luminosity is slightly below
the Eddington  luminosity  of a 14 M$_{\sun}$ star  and corresponds to a mass accretion rate 

%$$R_{circ} = 4(1+q)^{4/3}(0.500 - 0.227 LOG(q))^4P_{day}^{2/3},  \eqno(1)$$

$$\dot{M}_{obs} = 2.0 \times 10^{-7} (\eta/0.1)^{-1} (d/12.5 kpc)^2 M_{\sun}/year,
                                                           \eqno(3)$$

if $\eta$ is the  efficiency
of converting accretion to radiation 
(L = $\eta$$\dot{M}_{obs}$c$^2$). For a non-rotating black hole the maximum available
gravitational potential energy (per rest mass energy) at the innermost
 stable orbit 
is 0.06 -- 0.1, while for an extreme Kerr-hole the efficiency may be as high as 0.4 (see FKR, p. 191).

The observed high accretion rate eats the mass from the torus on a timescale 
t$_{active}$ =  M$_{accum}$/$\dot{M}_{obs}$.
We call this the   {\it 'activity time'}
 and it is  one order of magnitude shorter than the recurrence time 
= t$_{visc}$: 

 t$_{active}$/t$_{recurrence}$ = (0.05--0.1)($\eta$/0.1).
 
Together these two timescales  form
a {\it duty-cycle} and 
their estimates are presented in Table 2.

\section{Discussion and Conclusions}

We have estimated the He-core mass (around 0.28 M$_{\sun}$) of the donor of GRS 1915+105
for the binary parameters given by Greiner et al. (2001b).
The evolutionary expansion of the donor leads to a conservative mass transfer rate
$\dot{M}_d$ = (1.5 $\pm{0.5}$) $\times$ 10$^{-8}$ M$_{\sun}$/y which is ten times smaller than
the accretion rate  derived from the mean ASM light
curve over the past 6 years, and using  an efficiency of 
0.1 to convert the mass infall
into radiation: $\dot{M}_{obs}$ = 2.0$\times$($\eta$/0.1)$^{-1}$ 10$^{-7}$ M$_{\sun}$/y, for a distance
of 12.5 kpc.
We propose that these two numbers determine the duty cycle
where the active phase (as observed at present) is ten times shorter than the quiescent one. 

We identify the duration of the quiescent phase ({\it recurrence time}) as the viscous timescale at the circularization radius and estimate its value 
to be  370$\times$($\alpha$/0.001)$^{-4/5}$ years
($\alpha$ = the viscosity parameter in the $\alpha$-prescription).
 The corresponding active phase lasts 
28$\times$($\eta$/0.1)($\alpha$/0.001)$^{-4/5}$
  years 
and is comparable to the present activity phase which has already lasted for
 ten years, if the small $\alpha$ = 0.001 used can be justified at the outer edge of the disc. 

The 
$\alpha$-parameter
is highly uncertain and consequently so are 
the timescales derived. However, a comparison can be made with the 
recurrent X-ray Nova and Soft X-ray Transient A0620-003, using its parameters  during
 quiescence: M$_{BH}$ = 10 M$_{\sun}$,
M$_d$ = 0.7 M$_{\sun}$ and P = 7.75 hours 
(Tanaka and Lewin, 1995).  The accretion rate at the outer disc during quiescence, as 
derived from  optical
observations (McClintock et al. 1995), equals   $\dot{M}$$_d$ = 10$^{-10}$ M$_{\sun}$/y 
which  we identify as the mass transfer rate from the donor. 
We note that the  observed X-ray luminosity during quiescence implies much smaller accretion in the  inner regions of the disc (Narayan et al. 1996).
During the maximum outburst  the accretion onto the primary black hole probably approached  the Eddington rate   $\dot{M}$$_{obs}$= 10$^{-7}$  M$_{\sun}$/y for $\eta$ = 0.1, with an e-folding time of one month (Tanaka and Lewin, 1995). 
The circularization radius is  0.60 R$_{\sun}$ leading to t$_{visc}$ = 30 years
(using $\alpha$ = 0.001 in eq. 2) which is briefly consistent with 
the two observed outbursts  (1917 and
1975)  supporting the idea that the disc filling time is equal to the
viscosity timescale at R$_{circ}$ with small $\alpha$.
  The mass accumulated  is small (3 $\times$ 10$^{-9}$ M$_{\sun}$) 
and consequently the predicted active phase of A0620-003  is short (10 days) but of the same order of magnitude than the 
observed e-folding time.    
  
    At present there are more sophisticated disc-models available
than the simple $\alpha$-prescription used. These include e.g.
irradiated disc-models (see King, 2000, and references therein).  
Their usage would affect the viscosity timescale  for a fixed
$\alpha$ but probably less than the uncertainty in $\alpha$ itself.  

If the mass of the donor of GRS 1915+105 
is higher (2.4 M$_{\sun}$, instead of 1.2 M$_{\sun}$ 
as used in the above estimate),
the mass transfer from the donor will be increased to   
5$\times$10$^{-8}$ M$_{\sun}$/y. Further,
if at the same time we increase the efficiency of the BH conversion to radiation to 0.4 (instead of 0.1),
like in the case of an extreme prograde Kerr-hole, 
then $\dot{M}_d$ and $\dot{M}_{obs}$ become equal. In this most extreme case the reasons
behind the quiescent/active states must  be searched elsewhere, e.g. in strong advection
(ADAF) during the quiescence. 
We also checked that the hydrogen ionization zone (at around  R$_{\sun}$) 
is always inside the circularization radius  and may thus
be the trigger for the limit-cycle instability lasting 
for the whole activity phase.

The models in Table 1 
(including uncertainties in the donor mass) predict 
bolometric  luminosities 
L = (50 -- 100)L$_{\sun}$, surface effective
temperatures   T$_{eff}$ = 3800 -- 4000 K and  gravities 
 log(g) = 1.7 -- 1.9. These 
correspond to absolute 
K-magnitudes between $-$2.7 -- $-$2.2 which are inside the  limits
 ($-$2 -- $-$3) 
given by Greiner et al. (2001a), but
a more accurate value could  properly fix the donor mass, and consequently its
mass transfer rate. 

Another complication which should be studied is the possible effect of 
X-ray heating of the donor.
Hard photons above 10 keV can penetrate through its photosphere into the   
convective zone  affecting its structure
(Podsiadlowski, 1991;Vilhu, Ergma and Fedorova, 1994). The
mean luminosity of GRS 1915+105 above 10 keV is roughly 
2$\times$  10$^{38}$ erg/s (Vilhu et al. 2001)
of which 0.5 -- 1
per cent is captured by the donor, assuming no screening of the disc. If the activity phase
lasts 1/10 of the whole  cycle then 
(10--30)L$_{\sun}$ can be deposited in deep layers  of the donor,  
  averaged over the longer  thermal timescale of the donor envelope, 
leading probably to  an overestimate 
of the He-core mass and mass transfer rate.

\begin{acknowledgements}

I thank Diana Hannikainen and Ene Ergma 
for discussions and valuable comments and the anonymous referee for 
helping to make the paper clearer and to remove misprints. 

\end{acknowledgements}

\newpage

\vspace{3cm}
\begin{table*}
\caption{\label{tab:line}\protect\small 
Computed parameters for the evolving donor of GRS 1915+105
using M$_{BH}$ = 14 M$_{\sun}$,
M$_d$ = 1.2 M$_{\sun}$, P = 33.5 day  and assuming that the donor fills its 
Roche lobe (first line), using the analytic methods by Webbink, Rappaport
and Savonije (1983) for Z = 0.02. The 
second line gives the ranges 
 if the donor mass is varied between  (0.6 -- 2.4) M$_{\sun}$.
The He-core mass (M$_{He}$), luminosity (L), radius (R), binary separation (a)
 and the circularization radius (R$_{circ}$)
are given in solar units. The mass transfer from the donor ($\dot{M}$$_d$) is in units of 
solar masses per year. The predicted spectral types and absolute K-magnitudes (M$_K$)
are estimated from T$_{eff}$ and L using bolometric corrections and colours from
Cox (2000). }  
\small
\begin{center}
\begin{tabular}{ccccccccc}
\multicolumn{9}{c}{ } \\ 

Sp & M$_K$ & M$_{He}$  & L & R & $\dot{M}_d$ & a  & R$_{circ}$ &    \\ 
%\noalign{\smallskip}
%\hline
\small
%\vspace{0.5cm}

% \noalign{\smallskip}
K6 &$-$2.6 & 0.28 & 77 & 21 & 1.5 10$^{-8}$ & 108 & 14 &   \\

K5 -- M1 &$-$2.2 -- $-$2.7 &  0.26 -- 0.29 & 50 -- 100 & 17 -- 27 & 5 10$^{-9}$ -- 5 10$^{-8}$ & 95 -- 115 & 12 -- 18 &   \\

%\hline
\end{tabular}
\end{center}
\end{table*}
%\label{tabu1}
% $$ 
%  \begin{array}{p{0.5\linewidth}l}
%\begin{array}
%            \hline
%            \noalign{\smallskip}         
%Sp & M_K & M_{He}  & L & R & \dot{M}_d & a  & R_{circ} &    \\ 
%\noalign{\smallskip}
%\hline
%\noalign{\smallskip}
%K6 & $-$2.6 & 0.28 & 77 & 21 & 1.5 10^{-8} & 108 & 14   \\
%K5 -- M1 & $-$2.2 -- $-$2.7 &  0.26 -- 0.29 & 50 -- 100 & 17 -- 27 & 5 10^{-9} -- 5 10^{-8} & 95 -- 115 & 12 -- 18    \\
%   \noalign{\smallskip}
%            \hline
%         \end{array}
%     $$
%\centering
%\vspace{-9cm}
%\hspace{-4cm}
%\includegraphics[height=30cm]{donortable1.ps}
%\end{table*}    
%\\
\vspace{2cm}

\begin{table*}
\caption{\label{tab:line2}\protect\small
Viscosity time scale t$_{visc}$ (recurrence time) 
at the circularization radius, the mass M$_{accum}$
 accumulated from the donor during t$_{visc}$ and the time scale t$_{active}$ 
during which
the BH swallows  M$_{accum}$ with the observed mean accretion rate 
2$\times$10$^{-7}$($\eta$/0.1)$^{-1}$(d/12.5 kpc)$^2$ M$_{\sun}$/year
where $\eta$ is the accretion to radiation conversion factor and d the source
distance. 
{\it All the values should be multiplied   by ($\alpha$/0.001)$^{-4/3}$}
where $\alpha$ is the viscosity parameter 
in the $\alpha$-prescription of classical
discs. The models are 
the same as in Table 1.  }
\small
\begin{center}
\begin{tabular}{cccc}
\multicolumn{4}{c}{ } \\ 
  
t$_{visc}$ years & M$_{accum}$/M$_{sun}$ & t$_{active}$ / ($\eta$/0.1) years &   \\ 
%\hline

%\vspace{0.5cm}

370 & 5.5 $\times$ 10$^{-6}$  & 28 &  \\

200 -- 700 & (3.7 -- 9.0)$\times$ 10$^{-6}$  & 20 -- 45 &  \\

%\hline
\end{tabular}
\end{center}
\end{table*}

%\label{tabu2}           
%\small
%\begin{center}
%\centering
%             \hline
%\begin{tabular}{!R!R!R!}
%\multicolumn{3}{c}{ } \\ 
%    \hline
%\begin{array}{p{0.5\linewidth}l}
%\hline
%\noalign{\smallskip} 
%t$_{visc}$ years & M$_{accum}$/M$_{\sun}$ &  t$_{active}$ /  ($\eta$/0.1) years    \\
%\noalign{\smallskip} 
%\hline
%\noalign{\smallskip}
%370 & 5.5 $\times$ 10$^{-6}$  & 28    \\
%200 -- 700 & (3.7 -- 9.0)$\times$ 10$^{-6}$  & 20 -- 45    \\
%\noalign{\smallskip}
%\hline
%\end{tabular}
%\end{center}
%\end{array}
%\centering
%\vspace{-13cm}
%\vspace{-8cm}
%\hspace{-4cm}
%\includegraphics[height=30cm]{donortable2.ps}  
%\end{table*}

\end{document}